%% file: main.tex
\documentclass[a4paper, UKenglish, cleveref, autoref, thm-restate, pdfa]{lipics-v2021}
\nolinenumbers
\input{macro}

\ArticleNo{25}

\title{LISA -- A Modern Proof System}

\author{Simon Guilloud}{EPFL, Laboratory for Automated Reasoning and Analysis, Switzerland \and \url{https://people.epfl.ch/simon.guilloud} }{}{https://orcid.org/0000-0001-8179-7549}{}

\author{Sankalp Gambhir}{EPFL, Laboratory for Automated Reasoning and Analysis, Switzerland \and \url{https://people.epfl.ch/sankalp.gambhir} }{}{https://orcid.org/0000-0001-5994-1081}{}

\author{Viktor Kun\v{c}ak}{EPFL, Laboratory for Automated Reasoning and Analysis, Switzerland \and \url{https://lara.epfl.ch/~kuncak/} }{}{https://orcid.org/0000-0001-7044-9522}{}

\authorrunning{S. Guilloud, S. Gambhir and V. Kun\v{c}ak} 

\Copyright{Simon Guilloud, Sankalp Gambhir and Viktor Kun\v{c}ak} 



\date{2023}

\begin{document}
\maketitle

\begin{abstract}
    We present LISA, a proof system and proof assistant for constructing proofs in schematic first-order logic and axiomatic set theory.
    The logical kernel of the system is a proof checker for first-order logic with equality and schematic predicate and function symbols.
    It implements polynomial-time proof checking and uses the axioms of ortholattices (which implies the irrelevance of the order of conjuncts and disjuncts and additional propositional laws).
    The kernel supports the notion of theorems (whose proofs are not expanded), as well as definitions of predicate symbols and objects whose unique existence is proven.
    A domain-specific language enables construction of proofs and development of proof tactics with user-friendly tools and presentation, while remaining within the general-purpose language, Scala.
    We describe the LISA proof system and illustrate the flavour and the level of abstraction of proofs written in LISA. This includes a proof-generating tactic for propositional tautologies, leveraging the ortholattice properties to reduce the size of proofs. We also present early formalization of set theory in LISA, including Cantor's theorem.
\end{abstract}

\section{Introduction}

We present the design and initial implementation of a new proof assistant, named LISA. Much like Mizar \cite{naumowiczBriefOverviewMizar2009}, LISA aims to use classical mainstream foundations of mathematics with first order logic and set theory.
LISA uses (single-sorted) first-order logic (with schematic variables) as the syntactic framework, sequent calculus as the deduction framework and set theory as the semantic framework.  
On top of this foundation, we can construct mathematical theories without introducing additional axioms. As the target use of LISA we envision a library of theorems,  but also correctness proofs of computer systems.

LISA's source code and a reference manual, as well as all the examples in the present paper, are available from 

\centerline{\url{https://github.com/epfl-lara/lisa}}

\subsection{Design Goals}

Our design is inspired by the LCF line of proof assistants, including HOL Light, HOL4, and Isabelle. The envisioned path for axiomatic foundations is closer to Mizar. LISA's logical kernel is a hybrid between LCF-style encoding of theorems as a sealed Theorem type (similar to HOL Light \cite{harrisonHOLLightOverview2009}) and explicit requirement of proofs. Namely, proofs are self-contained sequences of proof steps that derive a conclusion from assumptions (they are not explicitly in the form of lambda terms). LISA's kernel checks the validity of steps and assumptions, and then creates an instance of a theorem. 

As the unified \emph{implementation, proof writing and tactic language}, we use Scala instead of the ML family of languages that are common to many proof assistants. Scala is a high-level functional and object-oriented language. We hope to avoid a sharp boundary between user proofs and tactic developments by using a single language with good support for domain-specific constructs. 
To provide a flavour of LISA, consider several ways of constructing proofs that are available to LISA users. \autoref{fig:exampleProofLinear} shows a proof of Pierce's law as an explicit sequence of sequent calculus proof steps. \autoref{fig:lisaCodeExample} and~\autoref{fig:lisaCodeExample2} show proofs built using a higher-level domain-specific language (DSL). This DSL detects high-level errors in incorrect proofs, but always generates the underlying lower-level proof and forwards it to the kernel to obtain a kernel-certified theorem. Finally,~\autoref{fig:OLSolver} shows a solver for propositional tautologies that uses the same mechanisms as the proofs to implement a proof tactic.
It was not our immediate goal to create an interactive experience, so our interaction model is more HOL4-like than Isabelle/HOL-like. For us, this means using Scala IDEs, rerunning projects, relying on incremental compilation. As the sizes of theories grow, we plan to develop serializations for proofs and theories to reduce re-execution.
We discuss both the kernel and the DSL in the rest of the paper.

The design philosophy of LISA focuses on what one might call the \emph{Six Virtues of Modern Proof Systems}. 
\emph{Efficiency} say a proof system components should have polynomial complexity (as close as possible to linear). 
\emph{Trust} means high confidence in the system, through a combination of well understood mathematical foundations,  explicit proofs and a concise logical kernel.
\emph{Usability} is making it simple, both for human users and automated methods, to formalize mathematics and  to develop tools. 
\emph{Predictability} is the property of systems whose behaviour and output have clear characterizations. 
\emph{Interoperability}, whose importance has become clear over the years, consists in making it as easy as possible for the system to be used by other systems and to export and import proofs to and from other systems. 
Finally, \emph{Programmability} implies that as a computer system, a proof assistant should provide all the expressiveness allowed by a programming language.
When designing and developing the LISA proof system, we aim to respect the six virtues as much as possible, and, when they oppose each other, to strike for the best balance between them.

\subsection{Contributions}

The contribution of this paper is to present the design of LISA, a new proof construction system embedded in Scala, based on schematic first-order logic with set theory axioms. We focus on the following aspects.
\begin{itemize}
    \item We describe how the logical kernel is constructed and how it can be used or interacted with by other tools.
    \item As the most unusual design aspect, we describe ortholattice-based algorithms implemented in the kernel to make proofs shorter.
    \item We present a domain-specific language embedded in Scala that makes the writing of proofs easier and generates and checks kernel proofs to obtain kernel-certified theorems.
    \item We show that the same domain-specific language can scale from writing proofs of specific theorems to writing general tactics. As an example of a tactic, we present a (proof generating) solver for propositional formulas leveraging the ortholattice algorithm.
    \item We report on the initial steps of developing elementary axiomatic set theory in the system.
\end{itemize}

\section{Logical Kernel}
LISA's deductive system is a variant of Gentzen's Sequent Calculus for first-order logic (FOL) \cite{Gentzen1935}. Formally, a sequent in LISA is a pair of sets of formulas $\Gamma$ and $\Delta$, represented
$
\Gamma \vdash \Delta
$
and its interpretation is $\bigwedge \Gamma \rightarrow \bigvee \Delta$.
LISA extends the prototypical Sequent Calculus with schematic symbols, substitution rules, and a normalization of formulas.

\subsection{Schematic Symbols}
Formulas in LISA's kernel are built with the usual variables, constant function and predicate symbols, logical connectors and binders, but also admit the use of schematic function, predicate and connector symbols. These symbols behave like uninterpreted constant symbols which can be substituted by any well-typed term or formula across a whole sequent, or like variables which cannot be bound.
We refer to them as second-order schematic symbols, as opposed to regular variables, which are first-order schematic symbols. This gives the system a flavour of second order logic and allows writing axiom and theorem schemas, as the following example illustrates:
\begin{example}
The following sequent (whose proof we show in \autoref{fig:lisaCodeExample}) is provable without additional assumptions on the function symbol $f$ and the predicate symbol $P$:
\[
\forall x. P(x) \rightarrow P(f(x)) \vdash \forall x. P(x) \rightarrow P(f(f(x)))
\]
This means that this sequent with f replaced by a specific term (with a distinguished free variable) remains provable by an analogous proof, and similarly for $P$ replaced by a formula.
\end{example}

In traditional first-order logic, this concept is formalized as a meta-theorem stating that for every $f$ and $P$, a corresponding proof can be built. However, in a formal setting, this requires duplicating the whole proof for every specific $f$ and $P$. Instantiation of schematic symbols avoids this issue of proof duplication.
This is similar to the schematic variables found in Isabelle \cite{DBLP:journals/corr/cs-LO-9301106}, and in particular {Isabelle/FOL} \cite{paulson2013isabelle}, where variables from the meta-logic can be used to represent arbitrary functions, predicates, and connectors in FOL.
Crucially, it does not increase the expressive power of the system, because it can, in principle, be simulated.

\subsection{Ortholattice Algorithm Applied to First-Order Logic}
\label{sec:OLalgo}

We find that using a proof system that is sensitive to the order of conjuncts and similar semantically irrelevant syntactic differences can be frustrating and increases proof size unnecessarily.
To address this issue, LISA's kernel strengthens sequent calculus with a built-in algorithm to compute normal form and equivalence of formulas with respect to a subset of equational rules of propositional logic. These rules, shown in \autoref{tab:laws}, characterize the algebraic theory of ortholattices (abbreviated OL)
\cite[Chapter II.1]{Beran8511ortholattices}, \cite{Bruns2000}.

Ortholattices are a generalization of Boolean algebra where instead of the law of distributivity, the weaker absorption law (L9, \autoref{tab:laws}) holds. In particular, every identity in the theory of ortholattices is also a theorem of propositional logic. 

\begin{table}[bth]
    \centering
    \begin{tabular}{r c @{\hskip 2em} | @{\hskip 2em} r c}
         L1: & $x \lor y = y \lor x$  & L1': & $x \land y = y \land x$ \\
         L2: & $x \lor ( y \lor z) = (x \lor y) \lor z$  & L2': & $x \land ( y \land z) = (x \land y) \land z$ \\
         L3: & $x \lor x = x$  & L3': & $x \land x = x$ \\
         L4: & $x \lor 1 = 1$  & L4': & $x \land 0 = 0$ \\
         L5: & $x \lor 0 = x$  & L5': & $x \land 1 = x$ \\
         L6: & $\neg \neg x = x$  & L6': & same as L6  \\
         L7: & $x \lor \neg x = 1$  & L7': & $x \land \neg x = 0$ \\
         L8: & $\neg (x \lor y) = \neg x \land \neg y$  & L8': &  $\neg (x \land y) = \neg x \lor \neg y$ \\
         L9: & $x \lor (x \land y) = x$ & L9': & $x \land (x \lor y) = x$   \\
    \end{tabular}
    
    \
    
    \caption{Laws of ortholattices, an algebraic theory with signature $(S, \land, \lor, 0, 1, \neg)$. \cite{Guilloud:297701}
    \label{tab:laws}}
\end{table}

This algebraic structure has been shown to possess a quadratic-time normalization algorithm \cite{Guilloud:297701} and has been suggested as the basis for normalization of formulas in the context of verification and mechanized proofs. Notably, it subsumes negation normal form.

As a special kind of lattices, ortholattices can be viewed as partially ordered sets, with the ordering relation on two elements $a$ and $b$ of an ortholattice defined as
\(
a\leq b \Longleftrightarrow a \land b = a
\), which, by absorption (L9), is also equivalent to $a \lor b = b$. If $s$ and $t$ are terms over the signature $(S, \land, \lor, 0, 1, \neg)$, we denote $s \leq_\OL t $ if and only if $OL \vDash s \leq t$, i.e., it holds in all ortholattices. 
We write $s\sim_\OL t$ if both $s\leq_\OL t$ and $s\geq_\OL t$ hold (or equivalently, if $OL \vDash s = t$).
\autoref{thm:OL} is the main result we rely on.

\begin{thm}[\cite{Guilloud:297701}]
\label{thm:OL}
There exists an algorithm running in worst case quadratic time producing, for any terms $s$ over the signature $(\land, \lor, \neg)$, a normal form $\text{NF}_{\OL}(s)$
such that for any $t$, $s \sim_\OL t$ if and only if $\text{NF}_{\OL}(s) = \text{NF}_{\OL}(t)$. The algorithm is also capable of deciding if $s \leq_{OL} t$ holds in quadratic time.
\end{thm}
Moreover, the algorithm works with structure sharing with the same complexity, which is very relevant for example when $x \leftrightarrow y$ is expanded to $(x \land y) \lor (\neg x \land \neg y)$. It can produce a normal form in this case as well.

These properties, along with completeness characterization, make the OL algorithm a good candidate to include in a proof system. LISA's kernel further extends OL inequality algorithm to first order logic formulas as follows. It first expresses the formula using de Bruijn indices \cite{de1972lambda}, then desugars $\exists. \phi$ into $\neg \forall. \neg \phi$. It then extends the OL algorithm with the rules in \autoref{tab:Olextension}.

\begin{table}[ht]
    \centering
    \begin{tabular}{c | l | l}
        & To decide... & Reduce to...\\
         \hline
       1&  $\lbrace \land, \lor, \rightarrow, \leftrightarrow, \neg \rbrace(\vec{\phi}) \leq \psi $ & Base algorithm \\
        2& $\phi \leq \lbrace \land, \lor, \rightarrow, \leftrightarrow, \neg \rbrace(\vec{\psi}) $ & Base algorithm \\
        3& $s_1 = s_2 \leq t_1 = t_2$ & $\lbrace s_1, s_2 \rbrace == \lbrace t_1, t_2 \rbrace$\\
        4& $\phi \leq t_1 = t_2$ & $t_1 == t_2$\\
        5 & $\forall. \phi \leq \forall. \psi$ & $\phi \leq \psi$\\
        6& $\schem{C}(\phi_1,...,\phi_n) \leq \schem{C}(\psi_1,...,\psi_n)$ & $\phi_i \sim_\OL \psi_i$, for every $1 \le i \le n$\\
         
        7& Anything else & \texttt{false}
    \end{tabular}
    
    \
    
    \caption{Extension of OL algorithm to first-order logic. We call it the \FOLalg{} algorithm. $=$ denotes the equality predicate in FOL, while $==$ denotes syntactic equality of terms.
    \label{tab:Olextension}}
\end{table}
When either of the two formulas being compared have a top-level propositional operator (cases 1 and 2), the recursion is done according to the algorithm described in \cite{Guilloud:297701}, considering any non-propositional expressions (predicates, quantified formulas, and schematic connectors) as propositional variables.
The third and fourth rules take into account reflexivity and symmetry of equality. The fifth relies on monotonicity of $\forall$, and the  sixth rule applies when $\schem{C}$ is a schematic connector, i.e., a logical connector about which we know nothing. These rules extend to the normal-form-producing algorithm, and it is easy to see that if $\leq$ is interpreted as logical implication, they are sound. 
We decided not to include a rule such as $\forall. \phi \leq \phi(t)$. The reason is that incorporating such a rule systematically runs risk of introducing higher complexity \cite{KapurNarendran92} in the kernel. We instead decided that such steps should be implemented using tactics, outside the kernel in the future (possibly making use of type-like hints encoded in first-order logic \cite{GanzingerMeyerWeidenbach97}). 

Using the First Order Logic OrthoLattices algorithm, noted \FOLalg{}, the proof checker in LISA's kernel performs every correctness check up to \FOLalg{} equivalence. This does not prevent sequents and formulas from having arbitrary constructions and being inspected in a stable, predictable way by tactics, as formulas are not normalized in-place. The set of LISA deduction rules is shown in \autoref{fig:deduct_rules_1}.

\begin{figure}
    \begin{center}
        \begin{tabular}{l l}
            \multicolumn{2}{c}{
                \AxiomC{}
                \RightLabel{\text { Hypothesis}}
                \UnaryInfC{$\Gamma, \phi \vdash \phi, \Delta$}
                \DisplayProof
            }\\[5ex]

            \multicolumn{2}{c}{
                \AxiomC{$\Gamma \vdash \phi, \Delta$}
                \AxiomC{$\Sigma, \phi \vdash \Pi$}
                \RightLabel{\text{ Cut}}
                \BinaryInfC{$\Gamma, \Sigma \vdash \Delta, \Pi$}
                \DisplayProof
            }\\[5ex]

            \AxiomC{$\Gamma, \phi, \psi \vdash \Delta$}
            \RightLabel{\text { LeftAnd}}
            \UnaryInfC{$\Gamma, \phi \land \psi \vdash \Delta$}
            \DisplayProof &
            \AxiomC{$\Gamma \vdash \phi, \Delta$}
            \AxiomC{$\Sigma \vdash \psi, \Pi$}
            \RightLabel{\text{ RightAnd}}
            \BinaryInfC{$\Gamma, \Sigma \vdash \phi \land \psi,  \Delta, \Pi$}
            \DisplayProof
            \\[5ex]

            \AxiomC{$\Gamma, \phi \vdash \Delta$}
            \AxiomC{$\Sigma, \psi \vdash \Pi$}
            \RightLabel{\text{ LeftOr}}
            \BinaryInfC{$\Gamma, \Sigma, \phi\lor \psi \vdash \Delta, \Pi$}
            \DisplayProof &
            \AxiomC{$\Gamma \vdash \phi, \psi \Delta$}
            \RightLabel{\text{ RightOr}}
            \UnaryInfC{$\Gamma \vdash \phi \lor \psi,  \Delta$}
            \DisplayProof
            \\[5ex]

            \AxiomC{$\Gamma \vdash \phi, \Delta$}
            \AxiomC{$\Sigma, \psi \vdash \Pi$}
            \RightLabel{\text{ LeftImplies}}
            \BinaryInfC{$\Gamma, \Sigma, \phi\rightarrow \psi \vdash \Delta, \Pi$}
            \DisplayProof &
            \AxiomC{$\Gamma, \phi \vdash \psi, \Delta$}
            \RightLabel{\text{ RightImplies}}
            \UnaryInfC{$\Gamma \vdash \phi \rightarrow \psi,  \Delta$}
            \DisplayProof
            \\[5ex]

            \AxiomC{$\Gamma, \phi \rightarrow \psi \vdash \Delta$}
            \RightLabel{\text { LeftIff}}
            \UnaryInfC{$\Gamma, \phi \leftrightarrow \psi \vdash \Delta$}
            \DisplayProof &
            \AxiomC{$\Gamma \vdash \phi \rightarrow \psi, \Delta$}
            \AxiomC{$\Sigma \vdash \psi \rightarrow \phi, \Pi$}
            \RightLabel{\text{ RightIff}}
            \BinaryInfC{$\Gamma, \Sigma \vdash \phi \leftrightarrow \psi,  \Delta, \Pi$}
            \DisplayProof
            \\[5ex]

            \AxiomC{$\Gamma \vdash \phi, \Delta$}
            \RightLabel{\text { LeftNot}}
            \UnaryInfC{$\Gamma, \neg \phi \vdash \Delta$}
            \DisplayProof &
            \AxiomC{$\Gamma, \phi \vdash \Delta$}
            \RightLabel{\text{ RightNot}}
            \UnaryInfC{$\Gamma \vdash \neg \phi ,  \Delta$}
            \DisplayProof
            \\[5ex]

            \AxiomC{$\Gamma, \phi[t := \schem{x}] \vdash \Delta$}
            \RightLabel{\text { LeftForall}}
            \UnaryInfC{$\Gamma, \forall \schem{x}. \phi  \vdash \Delta$}
            \DisplayProof &
            \AxiomC{$\Gamma \vdash \phi, \Delta$}
            \RightLabel{\text { RightForall}}
            \UnaryInfC{$\Gamma \vdash \forall \schem{x}. \phi,  \Delta$}
            \DisplayProof
            \\[5ex]

            \AxiomC{$\Gamma, \phi \vdash \Delta$}
            \RightLabel{\text { LeftExists}}
            \UnaryInfC{$\Gamma, \exists \schem{x}. \phi \vdash \Delta$}
            \DisplayProof &
            \AxiomC{$\Gamma \vdash \phi[t := \schem{x}], \Delta$}
            \RightLabel{\text { RightExists}}
            \UnaryInfC{$\Gamma \vdash \exists \schem{x}. \phi,  \Delta$}
            \DisplayProof
            \\[5ex]

            \multicolumn{2}{c}{
                \AxiomC{$\Gamma \vdash \Delta$}
                \RightLabel{\text{ InstSchema}}
                \UnaryInfC{$\Gamma[\psi(\vec{v}) := {\schem{p}(\vec{v})}] \vdash \Delta[\psi(\vec{v}) := {\schem{p}(\vec{v})}]$}
                \DisplayProof 
            }\\[5ex]

            \AxiomC{$\Gamma, \phi[s := \schem{f}] \vdash \Delta$}
            \RightLabel{\text{ LeftSubstEq}}
            \UnaryInfC{$\Gamma, s=t, \phi[t := \schem{f}] \vdash \Delta$}
            \DisplayProof &
            \AxiomC{$\Gamma \vdash \phi[s := \schem{f}], \Delta$}
            \RightLabel{\text{ RightSubstEq}}
            \UnaryInfC{$\Gamma, s=t \vdash \phi[t := \schem{f}], \Delta$}
            \DisplayProof
            \\[5ex]

            \AxiomC{$\Gamma, \phi[a := {\schem{p}}] \vdash \Delta$}
            \RightLabel{\text{ LeftSubstIff}}
            \UnaryInfC{$\Gamma, a \leftrightarrow b, \phi[b := {\schem{p}}] \vdash \Delta$}
            \DisplayProof &
            \AxiomC{$\Gamma \vdash \phi[a := {\schem{p}}], \Delta$}
            \RightLabel{\text{ RightSubstIff}}
            \UnaryInfC{$\Gamma, a \leftrightarrow b \vdash \phi[b := {\schem{p}}], \Delta$}
            \DisplayProof
            \\[5ex]

            \AxiomC{$\Gamma, t = t \vdash \Delta$}
            \RightLabel{\text { LeftRefl}}
            \UnaryInfC{$\Gamma \vdash \Delta$}
            \DisplayProof &
            \AxiomC{}
            \RightLabel{\text{ RightRefl}}
            \UnaryInfC{$\vdash t=t$}
            \DisplayProof
            \\[5ex]

            \multicolumn{2}{c}{
                \AxiomC{$\Gamma_1 \vdash \Delta_1$}
                \RightLabel{\text{ Restate} \text{ if $(\bigwedge\Gamma_1 \rightarrow \bigvee \Delta_1) \sim_\FOLm (\bigwedge\Gamma_2 \rightarrow \bigvee \Delta_2)$}}
                \UnaryInfC{$\Gamma_2 \vdash \Delta_2$}
                \DisplayProof 
            }\\[5ex]

            \multicolumn{2}{c}{
                \AxiomC{$\Gamma_1 \vdash \Delta_1$}
                \RightLabel{\text { Weakening} \text{ if $(\bigwedge\Gamma_1 \rightarrow \bigvee \Delta_1) \leq_\FOLm (\bigwedge\Gamma_2 \rightarrow \bigvee \Delta_2)$}}
                \UnaryInfC{$\Gamma_2 \vdash \Delta_2$}
                \DisplayProof
            }
        \end{tabular}
    \end{center}
\caption{Deduction rules allowed by LISA's kernel. Different occurrences of the same symbols need not represent equal elements, but only elements with the same \FOLalg{} normal form.}
\label{fig:deduct_rules_1}
\end{figure}

Moreover, the proof checker contains a special \texttt{Restate} proof step, which permits \FOLalg-transformations on the entire sequent, leveraging the interpretation of a sequent as a formula (an implication). We also leverage specifically the partial order computed by \FOLalg{} to expand the usual \texttt{Weakening} rule so that the premise sequent only has to be $\leq_\FOLm$ stronger than the conclusion, with both interpreted as formulas.
\texttt{Weakening} clearly subsumes \texttt{Restate}, but the latter ensures that the transformation is actually an equivalence and hence could be reversed, which can be a useful safeguard in practice. These rules subsume most propositional rules in \autoref{fig:deduct_rules_1}.

\subsection{Substitution Rules}

The substitution rules substitute equal terms or equivalent formulas inside a formula. They are deduced steps whose simulation from simpler steps can take a number of steps linear in the size of the sequent, yet are very frequent both in human-written proofs and automated reasoning (as done by SAT solvers or in systems with rewrite rules, for example), justifying their inclusion as base steps. A special case of substitution that is particularly important is the following: 

\begin{center}
    \AxiomC{$\phi \vdash \psi$}
    \RightLabel{\text { SubstIff}}
    \UnaryInfC{$\phi \vdash \psi[\phi := \top]$}
    \DisplayProof
\end{center}
This holds in a single step because $\phi \leftrightarrow \top \sim_\FOLm \phi$. In fact, \texttt{Restate} and \texttt{SubstIff} form a complete basis for propositional logic that we will leverage in \autoref{sec:tautology} to write a complete proof-producing tactic for propositional logic.

The inclusion of \FOLalg{} and the substitution and instantiation deduced rules in the logical kernel is a slight bend to the {trust} principle, but as the algorithm is only 300 lines of code, this is largely overshadowed by the increased usability and shorter proofs. In fact, the whole kernel adds up to a grand total of only 1607 lines of code. This comprises the implementation of first-order logic, the \FOLalg{} algorithm, first and second-order substitution, the sequent calculus steps, the proof checker, and a manager for definitions and theorems (detailed in \autoref{subs:theories}). Moreover, LISA's kernel is efficient: except for the quadratic \FOLalg{} algorithm, every procedure in the kernel is linear (up to logarithmic coefficients) in the size of the formulas or proofs being considered.

\subsection{Proof Objects} 
In LISA, a proof is an explicit list of proof steps, where each step can refer to previous steps via their respective position in the list and be referred by multiple subsequent steps. In other words, a proof is represented as a  topological linearization of the proof tree, or, more generally, a directed acyclic graph (permitting reuse of intermediate steps). A proof step also contains the arguments that allow the proof checker to efficiently verify it. In particular, LISA's kernel does not rely on a unification algorithm to check correctness of proof steps related to quantifiers.

Moreover, proofs are standalone objects checkable and exportable without the need for any kind of context. \autoref{fig:exampleProofLinear} shows an example of sequent calculus proof as a sequence of steps. Each step lists a sequent with a rule from \autoref{fig:deduct_rules_1} and a list of (the position of) previous steps from which the sequent follows.  \autoref{fig:ExampleProofLinearScala} shows executable Scala code that denotes the same proof, which can be given directly to the LISA kernel. The kernel can efficiently check its correctness and create a theorem whose statement corresponds to the last sequent, corresponding to the root of the proof tree. Note that, in this particular case, the same conclusion could be reached in a single step using the \texttt{Restate} rule.

If the proof relies on external theorems, axioms or definitions, those are stated after the list of proof steps and referred to with negative positions. We call those \emph{imported} sequents (\emph{imports}, for short). 
We adopt an analogous mechanism to support \emph{subproofs}. A subproof simulates deduced steps by encapsulating an inner proof and appears as a single step in the outer proof. In that case, the premises of the subproof become imports of the inner proof.

\begin{figure}[hbt]
\centering
\small
\begin{equation*}
    \begin{split}
    0 &\texttt { Hypothesis} &\quad \phi &\vdash \phi\\
    1 &\texttt { Weakening}(0) &\quad  \phi &\vdash \phi, \psi\\
    2 &\texttt { RightImplies}(1) &\quad  &\vdash \phi, (\phi \to \psi)\\
    3 &\texttt { LeftImplies}(2,0) &\quad (\phi \to \psi) \to \phi &\vdash \phi\\
    4 &\texttt { RightImplies}(3) &\quad &\vdash ((\phi \to \psi) \to \phi) \to \phi
    \end{split}
    \end{equation*}

    \caption{The proof of Pierce's Law as a sequence of steps using classical Sequent Calculus rules.}
    \label{fig:exampleProofLinear}

\end{figure}

\begin{figure}[hbt]
    \centering
    \begin{lstlisting}[print, language=scala, showspaces=true]
val PierceLawProof = SCProof(IndexedSeq(
    Hypothesis(               $\phi$ $\vdash$ $\phi$,                                 $\phi$),
    Weakening(                $\phi$ $\vdash$ ($\phi$, $\psi$),                            0),
    RightImplies(            () $\vdash$ ($\phi$, $\phi\implies\psi$),                1, $\phi$, $\psi$),
    LeftImplies( ($\phi\implies\psi$)$\implies\phi$ $\vdash$ $\phi$,                2, 0, ($\phi\implies\psi$), $\phi$),
    RightImplies(            () $\vdash$ (($\phi\implies\psi$)$\implies\phi$)$\implies\phi$,
                                                 3, ($\phi\implies\psi$)$\implies\phi$, $\phi$)
), Seq.empty /* no imports */ )
\end{lstlisting}
    \caption{The proof from~\autoref{fig:exampleProofLinear} written for LISA's kernel. $\vdash$ and $\implies$ are alternative, nicer constructors for sequents and formulas and are not part of the kernel. The second argument (here empty) is the sequence of proof imports.}
    \label{fig:ExampleProofLinearScala}
\end{figure}
\subsection{Theories}
\label{subs:theories}

LISA's proof checker can be used as a tool to produce and check proofs, independent of any context, but is not a sufficient tool to develop mathematical theories, as it lacks in particular the ability to make definitions. For this task, the kernel also offers a minimal utility to allow development of mathematical theories with the ability to introduce axioms, theorems, and definitions with guaranteed soundness.

\paragraph*{Theorems}
This part of the kernel, called the \lstinline{Theory}, is inspired from the LCF style \cite{Gordon1978EdinburghLA}. It allows checking a proof once, producing a value of a sealed type \lstinline{Theorem}, which can then be reused many times. The proof can then be forgotten. The \lstinline{Theory} will also verify that the given proof's imports are properly justified by existing axioms, theorems or definitions, so that the proven \lstinline{Theorem} can be considered unconditionally true, unlike its standalone proof. \autoref{fig:usingTheory} shows how to use the \lstinline{Theory} to obtain a theorem.

\begin{figure}[ht]
    \centering
    \begin{lstlisting}[language=scala]
val theory = new Theory
val pierceThm: theory.Theorem = theory.makeTheorem(
    "Pierce's Law",
    () $\vdash$ (($\phi\implies\psi$)$\implies\phi$)$\implies\phi$,
    PierceLawProof, 
    Seq.empty
)
    \end{lstlisting}
    \caption{The proof from \autoref{fig:ExampleProofLinearScala} can be transformed into a \lstinline{Theorem} by a \lstinline{Theory}. The arguments are, in order, the name of the theorem, its statement, a proof of the statement and the list of previous theorems, axioms or definitions used to justify the proof's imports, if any.}
    \label{fig:usingTheory}
\end{figure}

The \lstinline{Theory} naturally corresponds to the concept of a ``mathematical theory'' in first-order logic, containing the language and axioms of said theory. To allow coexistence of multiple different theories with different valid theorems, LISA makes the \lstinline{Theory} a class that can be instantiated multiple times. The \lstinline{Theorem} type is dependent on a specific instance of \lstinline{Theory}, so that two different theories will reject the theorem of the other. In a language without dependent types, this could be replaced by a simple runtime check. Note that in proof development, it is expected that the user will never need to use more than one theory at once, so this aspect is abstracted by the DSL.

 \paragraph*{Definitions}
The theory also allows introducing new definitions for predicate and function symbols.

A \emph{predicate symbol} $P$ definition is of the form $P(x_1, \ldots, x_n) := \phi_{x_1,\ldots,x_n}$,
where  the $x_1,\ldots,x_n$ are the free variables of a given formula $\phi$.
To define a \emph{function symbol} $f$, the definition requires a proof of unique existence of the form:
\begin{equation}
 \exists ! y.\ \phi_{y, x_1, \ldots, x_n}
\label{eqn:defUnique}
\end{equation}
and introduces a definitional axiom
\(
\phi_{f(x_1, \ldots, x_n),x_1,\ldots,x_n},
\)
where again $x_1,\ldots,x_n$ are the free variables of the formula $\phi$. To make such a definition, the \lstinline{Theory} checks that the symbol has not already been defined and requires a proof of~\eqref{eqn:defUnique}, i.e., of existence and uniqueness.

\smartparagraph{Remark on unique existence}
One may hope that only the existence (but not uniqueness) was needed to obtain conservative extensions in first-order logic. Unfortunately, this is not true in the presence of axiom schemas. In particular, with such a definition principle, it becomes possible to prove the Axiom of Choice in ZF set theory, while they are well known to be independent \cite{10.2307/71858}. Indeed, in ZF it is possible to prove
\[
\forall x. \exists y.\ (x \neq \emptyset \implies  y \in x)
\]
from which we would obtain a function $\pick$ with the property
\[
\forall x.\ (x\neq \emptyset \implies \pick(x) \in x)
\]
If then the symbol $\pick$ is allowed in axiom schemas, as would be the case in LISA, it is then easy to use $\pick$ and the replacement schema to construct a choice function on any set (see also LISA's Reference Manual \cite{guilloudLISAReferenceManual2023}).

\smartparagraph{Abstraction via underspecified definitions}
We have seen that we need uniqueness to ensure conservative extensions. On the other hand, such requirement often forces the defining formula
to be overly specific and representation-dependant.
For example, one may want to define the set of real numbers, $\mathbb{R}$, as a structure that satisfies the axioms of real closed fields. Since there are many isomorphic structures satisfying these, a uniqueness proof cannot be obtained. It is then necessary to use a specific construction, such as Cauchy sequences, as the definition of the set of real numbers.
This, however, means that it becomes possible to prove properties of real numbers which are specific to the chosen representation, which is undesirable and especially so when transferring proofs to other proof systems, which may have different representations of reals. Our solution is to allow \emph{underspecified} definitions. An underspecified definition still requires existence and uniqueness (ensuring a conservative extension), but the theorem that the kernel provides is only the desired, weaker, one. This mechanism makes use of the $\leq$ relation of \FOLalg{}.
\autoref{sec:dsllisa} shows an example of the use of underspecified definitions in set theory.

This issue is addressed in Metamath \cite{megill2006metamath} by assuming a specific construction of the structure to conditionally prove a desired defining property (for example, the axioms of the real field) and then introducing said property independently as an axiom. This mechanism however is not enforced by Metamath itself but only an informal practice. LISA's kernel support for underspecified definitions ensures that the same goal is achieved with guaranteed soundness. Underspecification was also discussed in the context of HOL by Rob Arthan \cite{arthanHOLConstantDefinition2014}. Our approach is similar but the challenges are different. The use of Hilbert's description operator leads to undesired properties being provable, similarly as definition via unique existence, but for the reason explained above, Lisa can't relax the condition to existence only due to the presence of axiom schemas. Forgetting a part of the definition after it was made was tried in HOL Light, but this made reasoning about the system harder, as it has to take into account not only the state of the system but also the sequence of operation leading to this state. LISA avoid this issue by making underspecified definitions an integral part of the foundation.

\section{DSL for LISA in Scala}
\label{sec:dsllisa}

While the minimality of the kernel makes it tedious to use directly, the tools offered by Scala (and especially Scala 3) allow us to design a more intuitive DSL, similar to other proof assistants, directly within the host language. Moreover, essentially all the verification related to the syntactic construction and writing of the proof are checked at compilation time, leaving only the wrong use of proof steps and tactics (such as when trying to prove an invalid statement) as possible failure at runtime. LISA's interface encapsulates the kernel and provides convenient tools and syntax to make mathematical development easier to write and read. \autoref{fig:lisaCodeExample} shows a minimal example of how to use the DSL to write a proof. This approach makes LISA programmable. It offers the user the full range of tools of the host language when writing proofs, allowing them to express proofs in novel ways or adapted to different areas of mathematics, similar to writing on paper.

LISA's environment is activated simply by creating an object extending \lstinline{lisa.Main}. This will make available all the essential features to develop mathematics in LISA. Declarations in lines 2, 3 and 4 define a variable, a schematic predicate of arity one, and a schematic function of arity one, such that their symbols are the same as their Scala name, i.e., respectively \lstinline{"x"}, \lstinline{"P"} and \lstinline{"f"}. This is made possible by implicit arguments and reflection. Line 6 starts the declaration of a theorem. (Note that the kernel itself does not rely on such specific features; we expect the kernel to be straightforward to implement in most languages.)

    

\begin{figure}[ht]
    \centering
        \begin{lstlisting}[print, language=lisa]
object Exercise extends lisa.Main {
    val x = variable
    val P = predicate(1)
    val f = function(1)
    
    val fixedPointDoubleApplication = Theorem(
            $\forall$(x, P(x)$\implies$P(f(x))) $\vdash$ P(x)$\implies$P(f(f(x)))
        ) {
        assume($\forall$(x, P(x)$\implies$P(f(x))))
        val step1 = have(P(x)$\implies$P(f(x))) by InstantiateForall
        val step2 = have(P(f(x))$\implies$P(f(f(x)))) by InstantiateForall
        have(thesis) by Tautology.from(step1, step2)
    }
}
        \end{lstlisting}  
    \caption{A small proof written with LISA's DSL. Unicode characters are obtained in practice through ligatures or Scala's direct support for unicode.
    }
    \label{fig:lisaCodeExample}
\end{figure}

\subsection{Higher-Level Proofs}

LISA's interface defines a proof constructing class. This class uses proof tactics to generate pieces of the final pure sequent calculus proof, which are encapsulated into kernel subproofs. The result from the point of view of the user is the ability to define arbitrarily computed deduced proof steps (here \lstinline{Tautology} and \lstinline{InstantiateForall}) from the base steps of sequent calculus. 
Thanks to Scala 3's implicit functions types \cite{Odersky:229203}, the proof constructor is automatically created in the code block following the \lstinline{Theorem} declaration (line 6 of \autoref{fig:lisaCodeExample}) without the need for the user to even realize it exists. The existence of an implicit proof constructor in scope is necessary for the other keywords (\lstinline{have}, \lstinline{assume}, ...) to be well-defined, meaning that using those outside of a theorem environment will fail to compile.

The \lstinline{assume} keyword (line 9) allows stating a formula that will be assumed true for the rest of the proof. Technically, it will be considered as part of the left-hand-side of any further written sequent in the proof.
\lstinline{have} states a proposition that can be reached using a proof tactic (or a subproof, see next example). If a step requires some premises, they can be given as parameters to the tactic, as in line 12. \lstinline{have} produces a \lstinline{Fact} that can be used by later steps.

\begin{figure}[ht]
    \centering
            \begin{lstlisting}[print, language=lisa]
val unionOfSingleton = Theorem( (union(singleton(x)) $\equiv$ x) ) {
    val X = singleton(x)
    val forward = have( (in(z, x)$\implies$in(z, union(X))) ) subproof {
        ...
    }    
    val backward = have( in(z, union(X))$\implies$in(z, x) ) subproof {
        have(in(z, y) $\vdash$ in(z, y)) by Restate
        val step2 = thenHave((y$\equiv$x, in(z, y)) $\vdash$ in(z, x)) 
            by Substitution
        have(in(z, y) $\land$ in(y, X) $\vdash$ in(z, x)) 
            by Tautology.from(pairAxiom of (y$\rightarrow$x, z$\rightarrow$y), step2)
        val step4 = thenHave($\exists$(y, in(z, y) $\land$ in(y, X)) $\vdash$ in(z, x)) 
            by LeftExists
        have( in(z, union(X))$\implies$in(z, x)) 
            by Tautology.from(unionAxiom of (x $\rightarrow$ X), step4)
    }    
    have( in(z, union(X)) $\iff$ in(z, x)) 
        by RightIff(forward, backward)
    thenHave( forall(z, in(z, union(X)) $\iff$ in(z, x))) 
        by RightForall
    andThen(Substitution(extensionalityAxiom of (x $\rightarrow$ union(X), y $\rightarrow$ x)))
}
        \end{lstlisting}
    \caption{A LISA proof with more advanced construction.}
    \label{fig:lisaCodeExample2}
\end{figure}

The example in \autoref{fig:lisaCodeExample2} illustrates a more advanced proof structure, using axioms from set theory. As the proof independently proves both directions of the double implication, it makes use of the \lstinline{subproof} construction. Similarly to the \lstinline{Theorem} keyword, this construction implicitly creates a new proof constructor environment, internal to the outer proof and with its own goal.
In a proof, a \lstinline{Fact} is a type that contains external theorems, axioms and definitions, as well as previously proven steps from the current or outer proof, but not from any proof that is not a direct ancestor of the current proof. This is made possible by using recursively defined path-dependent types (see \autoref{fig:factType}) and can be checked at compile-time.

Moreover, a fact can also be one of the above, accompanied by information about a specific instantiation of schematic symbols. The actual instantiation step is then carried automatically. This is done in practice with the \lstinline{of} keyword, as in line 11.

When a tactic requires a single premise, and this premise is the most recently proven fact, \lstinline{thenHave} passes said premise directly to the tactic without the step having to be named.
For some tactics, such as the \lstinline{Substitution} step at line 21, the resulting sequent will be inferred by the tactic and isn't required to be given by the user. In this case, \lstinline{have} and \lstinline{thenHave} takes the tactic as argument. The \lstinline{Tautology} step proves statements using propositional laws and the \lstinline{Substitution} makes substitution of equals for equals, either everywhere or using unification to find the specific occurrences to replace.

\begin{figure}[ht]
\begin{lstlisting}[print, language=scala]
class Proof {
    class ProofStep {...}
    class InnerProof extends Proof {
        val parent:Proof.this.type = Proof.this // The encapsulating proof
        type Fact = parent.Fact | this.ProofStep
    }
}
class BaseProof extends Proof {
    type Fact = Theorem | Axiom | Definition | this.ProofStep
}

\end{lstlisting}
\caption{Simplified outline of the type structure for proof constructors and their facts.}
\label{fig:factType}
\end{figure}

\subparagraph*{Definitions}
Transparent definitions come for free with the Scala host language (see line 2 of \autoref{fig:lisaCodeExample2}), these are not visible to the kernel. The DSL offers syntax for the \emph{non-transparent definitions}. Predicate symbol and function symbol (of which constant symbols are a special type) definitions can be direct, as illustrated by the two first examples in \autoref{fig:definitionExample}.
Function symbols can also be defined by unique existence, as shown in the last example. Note that this is an example of an \emph{underspecified definition}, as mentioned in the previous Section. It defines a constant symbol \lstinline{nonEmpty} with only the property \lstinline{$\neg$(nonEmpty$\equiv$$\emptyset$)}, but the given proof shows the existence of a specific non-empty set.

\begin{figure}[ht]
    \centering
        \begin{lstlisting}[print, language=lisa]
val succ = DEF(x) $\rightarrow$ union(uPair(x, singleton(x)))
val inductive = DEF(x) $\rightarrow$ in($\emptyset$, x) $\land$ $\forall$(y, in(y, x)$\implies$in(succ(y), x))
val nonEmptySetExists = Lemma( $\exists$!(x, $\neg$(x $\equiv$ $\emptyset$) $\land$ (x $\equiv$ uPair($\emptyset$, $\emptyset$)))){...}
val nonEmpty = DEF() $\rightarrow$ The(x, $\neg$(x $\equiv$ $\emptyset$))(nonEmptySetExists)
        \end{lstlisting}
    \caption{Definitions in LISA}
    \label{fig:definitionExample}
\end{figure}

\section{Tactics in LISA and Comparison}
Developing proof tactics in proof assistants where the proof-writing language is different from the host language (and sometimes when both are different from the tactic-writing language) tends to exhibit high entry barriers for newcomers. They require learning multiple new languages and how they interact with each other. This difficulty can be observed for example with the length of the tactic-writing tutorial for Isabelle \cite{urbanIsabelleCookbook2013}, or in the Coq Reference Manual, where the Ltac tactic language \cite{delahaye2000tactic} is described as \emph{having unclear semantics, being slow, non-uniform, error-prone} and even lacking essential programming features such as \emph{data structures}. Ltac2 \cite{Ltac2Coq16}, yet another tactic language, aims to solve \emph{some} of these problems. Newly developed systems, such as Lean \cite{demouraLeanTheoremProver2015}, have the advantage of being designed from scratch and addressing these problems. We have similar aims with LISA, but rely on an existing programming language which has already solved those issues, has an active user base that draws on more than the development of theorems and has well-developed and actively maintained IDEs and libraries. In particular, for a LISA user, seeing how a proof tactic works is ever only a ctrl-click away from their proof and when a new tactic is written, using it is as simple as writing \lstinline{import MyTactic}. 

Not unlike in HOL Light, where a proof tactic is essentially any function returning a value of type Theorem, a tactic in LISA is simply a function returning a proof or an error message.
The tactic can take arbitrary arguments, such as a target sequent and known facts (which will be imports of the resulting proof) and can access the current state of the proof constructor (if needed).
To write a low level or highly optimized proof tactic, the user can directly construct a sequent calculus proof and give it to the kernel, but they can also use LISA's DSL directly inside the body of the function and use pre-existing proof tactics. Writing a tactic then consists in writing a generic LISA proof computationally.

LISA defines tactics that correspond to each basic proof step within the kernel, but with all the parameters automatically inferred. These tactics are intended for didactic purpose. Compared to directly using kernel proof steps, these simple tactics are more convenient to write, but also slightly less efficient to check because the system needs to compute the parameters of the proof step. Moreover, most of these simple tactics are subsumed by more general tactics. 

\subsection{A Proof-Producing SAT Solver Using \texorpdfstring{\FOLalg{}}{FOL2}}
\label{sec:tautology}

The Tautology tactic is able to prove any valid sequent that requires only propositional reasoning. It is based on a simple proof-producing DPLL-like \cite{davisMachineProgramTheoremproving1962} procedure complete for propositional logic. The procedure makes decisions on atoms, so the worst case complexity is exponential in the number of unique atoms in the formula. It is a non-clausal solver (like, e.g., \cite{JainBartzisClarke06nonclausal}) whose unique aspect is that, between each decision, it simplifies the propositional formula using the algorithm presented in \autoref{sec:OLalgo}).
In the context of proving validity as in LISA as well as when trying to find a satisfying assignment in a SAT solver, this allows to close branches early in the exploration of the decision tree, or simply to eliminate atoms before they even need to get decided. Moreover, this procedure does not need to compute Tseytin's normal form, avoiding creating more atoms, and conveniently allows producing a proof of the statement. \autoref{fig:OLSolver} sketches the proof search procedure as it is implemented in LISA. Our current implementation uses a simple decision heuristics that picks the atom that occurs most frequently. Further work may also include extension of the algorithm with quantifier reasoning, to obtain a complete procedure for FOL.
\begin{figure}[ht]
    \centering
    \begin{lstlisting}[print, language=lisa]
def solveFormula(f: Formula,
                 decisionsPos:List[Formula],
                 decisionsNeg:List[Formula]): ProofTacticJudgement = {
  val redF = reduceWithFol2(f)
  if (redF == $\top$) {
    Restate(decisionsPos  $\vdash$ f :: decisionsNeg)
  } else if (redF == $\bot$) {
    InvalidProofTactic("Sequent is not a propositional tautology")
  } else {
    val atom = findBestAtom(redF)
    val substInRedF: Formula => Formula = (f => RedF[atom:=f])
    TacticSubproof {
      have(solveFormula(substInRedF($\top$), atom::decisionsPos, decisionsNeg))
      val step2 = thenHave(atom :: decisionsPos  $\vdash$ redF :: decisionsNeg)
              by Substitution($\top$ <=> atom)
      have(solveFormula(substInRedF($\bot$), decisionsPos, atom::decisionsNeg))
      val step4 = thenHave(decisionsPos  $\vdash$ redF :: atom :: decisionsNeg) 
              by Substitution($\bot$ <=> atom)
      thenHave(decisionsPos  $\vdash$ redF :: decisionsNeg) 
              by Cut(step2, step4)
      thenHave(decisionsPos  $\vdash$ f :: decisionsNeg) 
              by Restate
    }
  }
}
    \end{lstlisting}
    \caption{Outline of the \FOLalg{}-based solver. Note that the actual implementation produces directly kernel proofs for optimization. Each recursive call to \lstinline{solveFormula} adds at most 4 kernel steps to the final proof.}
    \label{fig:OLSolver}
\end{figure}

Thanks to properties of ortholattices, the solver is already capable of resolving propositional problems that are too difficult for some proof assistants. As an example, we found that Isabelle's Blast tactic (a general tableau prover, \cite{paulsonGenericTableauProver}) was in general not able to prove the equivalence of two reasonably large formulas made only of variables, disjunctions and conjunctions which only differed in the ordering of their arguments. On the other hand, this is instantaneous (one step) with our described OL-based approach. 

\subsection{Error Reporting}
LISA's DSL also contains a printer for proof (both kernel and high level) and defines specialized error reporting. Tactics are allowed to fail if they are used incorrectly and return an error. \autoref{fig:errorReporting} shows LISA's output for an incorrect proof, with the current state of the proof, the faulty step, its line number and the error message from the tactic.

\begin{figure}[ht]
    \centering
    \begin{lstlisting}[deletekeywords={in}]
!*\color{red}{$\forall$x. P(x)$\implies$P(f(x)) $\vdash$ P(x)$\implies$P(f(f(x)))}*!
  0 Hypothesis        $\forall$x. P(x)$\implies$P(f(x)) $\vdash$ $\forall$x. P(x)$\implies$P(f(x))
  1 Hypothesis        P(x); $\forall$x. P(x)$\implies$P(f(x)) $\vdash$ P(x)
  2 InstantiateForall P(x); $\forall$x. P(x)$\implies$P(f(x)) $\vdash$ P(x)$\implies$P(f(x))
  3 InstantiateForall P(x); $\forall$x. P(x)$\implies$P(f(x)) $\vdash$ P(f(x))$\implies$P(f(f(x)))
  !*\color{red}{have(thesis) by Tautology.from(step1)}*!

Proof tactic Tautology used in (Example.scala:47) did not succeed:
   The statement is not provable within propositional logic.
   The proof search needs the truth of the following sequent:
   P(f(x)); P(x); $\forall$x. $\neg$(P(x) $\land$ $\neg$P(f(x))) $\vdash$  P(f(f(x)))

    \end{lstlisting}
    \caption{LISA's output when the step in line 12 of proof in \autoref{fig:lisaCodeExample} is incorrectly modified to not use \lstinline{step2}. The indicated sequent in fact corresponds to \lstinline{step2}.}
    \label{fig:errorReporting}
\end{figure}

\section{Beginning Set Theory Development and Cantor's theorem}

In this section, we present a brief overview of the current mathematical development in LISA and outline an example of a short proof in set theory.

Inspired by 
Mizar\cite{naumowiczBriefOverviewMizar2009} and
Isabelle/HOTG\cite{brownHigherOrderTarskiGrothendieck2019} 
we make the choice of Tarski-Grothendieck set theory (\tg) as the axiomatic foundation for LISA's associated mathematical library. As the main reference for the \zfc aspect of the set theory development, we follow Thomas Jech's book \emph{Set Theory} \cite{jech2003set}. In the future, we plan to use the axiom on Grothendieck universes (corresponding to the existence of certain large cardinals) to support the embedding of category theory and of systems such as Coq \cite{wernerSetsTypesTypes1997}.


\subsection{Current Theory Development}

The mathematical library in LISA begins with the \zfc (and \tg) axioms, defining the basic constructs and operations on sets, the subset relation, the empty set, power sets, and unordered pairs. 
On top of thes axioms, we define structures such as ordered pairs, relations, and functions. Relations are sets of ordered pairs drawing elements from a set, and functions are relations which contain the graph of the function. Function symbols have as a domain the whole set space and must not be mixed with function objects, which are special sets and considered as constants in the light of first order logic. During exploratory development, proofs involving case analysis on these basic structures required significant manual effort, but the \lstinline{Tautology} and \lstinline{Substitution} tactics as well as the quick instantiation of axioms and theorems offered by the \lstinline{of} keyword tend to automate most tedious manipulations.
Formalization of partial orders, well-ordered sets, ordinals and induction \cite[Chapters 2, 3]{jech2003set} is ongoing. 

Technically, we define for sets \(A\) and \(B\) the set of relations from \(A\) to \(B\) as the power set of their Cartesian product \(P(A \times B)\), and its restriction to functional relations, \(A \to B\), the set of all functions with  domain equal to \(A\) and their codomain included in \(B\).

A function symbol can always apply to any term, meaning we cannot rely on well-definedness of terms to define symbols with partial function semantics. Considering the unique existence requirement for definitions, the standard approach consists in extending the limited domain of the partial function by assigning a default value, for example the empty set, to all inputs where it should be undefined, constructing a unique object. This specific construction and default value can then be forgotten using an underspecified definition. In particular, interpretations of the function with all combinations of values outside the fixed domain will be valid models for the symbol, and no non-trivial property can be proved about those values.

For example, consider the definition for function application, app\((f, x)\).
When \(f\) is not a functional relation, or \(x\) is not in its domain, we fix $\emptyset$ as the default value in order to obtain a proof of existence and uniqueness.
    \begin{lstlisting}[language=lisa]
val appDefinition = Theorem( $\forall$(f, ($\forall$ x, ($\exists$!(z, 
        functional(f) $\land$ in(x, dom(f)) $\implies$ in(pair(x, z), f)))) )
    $\land~ \neg$functional(f) $\lor~ \neg$in(x, dom(f)) $\implies$ z $\equiv \emptyset$)
    \end{lstlisting}
We can then obtain the function symbol \lstinline{app} with only the desired property using an underspecified definition:
    \begin{lstlisting}[language=lisa]
val app = DEF (f, x) $\rightarrow$ The(z, 
            functional(f) $\land$ in(x, dom(f))$\implies$in(pair(x, z), f))
            (appDefinition)
    \end{lstlisting}

\paragraph*{Cantor's Theorem}
Finally, several of these definitions and lemmas build up to the formalization of Cantor's theorem, stating that there is no surjection from any set to its power set:
    \begin{lstlisting}[language=lisa]
val cantorTheorem = Theorem( $\neg$surjective(f, x, powerSet(x)) )
    \end{lstlisting}
%
where \(f\) and \(x\) are schematic set variables, making the sequent implicitly universally quantified. The proof of Cantor's theorem is about 25 lines of code \footnote{\url{https://github.com/epfl-lara/lisa/blob/fc37f2a6e879d5f43679a4476c1d6e4685bb14a2/src/main/scala/lisa/mathematics/SetTheory.scala\#L1700}}. Internally, the proof expands to 130 sequent calculus steps.

Cantor's theorem is the first theorem formalized in LISA from the list \emph{Formalizing 100 Theorems} \cite{formalizing100}. While not a difficult theorem, it requires some ground development and definitions related to set-theoretic functions and relations. The proof itself requires handling the quantifiers for a contradiction construction and combining lemmas about surjective functions. Much of the latter is achieved using \texttt{Tautology}. It shows that LISA is capable of non-trivial mathematical development. We expect future developments to become easier and faster with gradual development of reasoning tools and proofs.



\section{Related Work}

A polynomial algorithm for free ortholattices was presented in \cite{Guilloud:297701}. A weaker structure with log linear complexity was first presented in \cite{guilloudEquivalenceCheckingOrthocomplemented2022}. In LISA we use the ortholattice normal form for first-order logic formulas.
Our \FOLalg{} implementation does not aim to be complete for structures such as quantum monadic algebras that treat extensions of OL (and orthomodular lattices) to monadic first-order logic \cite{Harding2022QuantumMonadic}.

Much of what we described is concerned with the schematic first-order logic kernel. We chose to include schematic variables to be able to state explicitly the axiom schemas of Zermelo-Fraenkel set theory and its extensions, as well as theorem schemas. Another way to generalize schematic second-order variables would be to use higher-order logic. This is the approach pursued by Isabelle as a framework, and instantiated in Isabelle/ZF.

The choice of set theory may be considered unusual by some, as Coq \cite{bertotInteractiveTheoremProving2004}, Lean \cite{demouraLeanTheoremProver2015}, the HOL-family \cite{harrisonHOLLightOverview2009} and Isabelle/HOL \cite{wenzelIsabelleFramework2008}  are based either on type theory or on higher-order logic.  We consider HOL to be one of the most elegant formulations for formal proof developments.
However, set theory is arguably the most widely recognized foundation of mathematics in the mathematical community, and, despite type-theory based tools having the advantage of being easier to express formalisms in from the get-go, we believe that through the development and use of abstracting tactics, a soft-type system and adequate tools, more familiarity and flexibility in writing proofs can be achieved with a set-theory based mathematical library. We also hope to provide a test bed to explore direct first-order foundations as an alternative to the many current systems based on higher-order logic.
Concrete results in Mizar \cite{naumowiczBriefOverviewMizar2009}, Isabelle/ZF \cite{Independence_CH-AFP}, ZF in Isabelle/HOL \cite{brownHigherOrderTarskiGrothendieck2019, ZFC_in_HOL-AFP}, and TLA$^+$ \cite{10.1007/978-3-642-32759-9_14,tlaps} suggest substantial relevance of set-theoretic foundations. Arguments in favour of set theoretic foundations have also been discussed by John Harrison \cite{harrisonLetMakeSet} and Bohua Zhan \cite{zhanFormalizationFundamentalGroup2017}.

Even one more level of indirection than in Isabelle/ZF is present in Isabelle/HOL/TG \cite{brownHigherOrderTarskiGrothendieck2019}, which develops the Tarski-Grothendieck extension of \zf inside Isabelle/HOL. Whereas our system is less flexible and does not currently connect to such a well-developed ecosystem as Isabelle/HOL has, our hope is that it is conceptually simpler thanks to fewer layers and a kernel that does not rely on unification.

Another modern approach to theorem proving is Lean \cite{demouraLeanTheoremProver2015}, a proof assistant based on dependent type theory and inspired in part by Coq. We believe Lean makes significant improvements over older proof assistant regarding the \emph{Six Virtues}. In particular, it has a strong focus on programmability, with the new version of Lean \cite{demouralean4} even having a compiler written in its own proof language. While LISA and Lean's design objectives share similarities, their strategies and specific choice (foundations, language, interface) are different.

To automate proofs that do not instantiate schematic formulas we hope to make use of proof generating theorem provers, 
such as Vampire \cite{kovacsFirstOrderTheoremProving2013}, SPASS \cite{weidenbachSPASSVersion2009}, E \cite{schulzSystemDescription2013}, as well as 
 SMT solvers \cite{barbosaCvc5VersatileIndustrialStrength2022}.
 Higher-order provers such as
 Zipperposition \cite{vukmirovicMakingHigherOrderSuperposition2021}, Leo-III \cite{alexander_steen_2022_7650205}, and Satallax \cite{10.1007/978-3-642-31365-3_11} would further increase automation even in the case of axiom schema instantiation.

\section{Conclusion}
LISA is both a proof system for automated tools and a proof assistant based on first order logic and set theory. It uses Scala as both a host language and a proof writing language, relying on the advanced features it offers to make the system as programmable as the user desires. LISA is strongly committed to interoperability. In particular, it has a small logical kernel which has guaranteed complexity and completeness characterizations, simple foundations and explicit proofs checkable without context. Moreover, it can be compiled into a Scala and Java library. All these properties should favour transfer of proofs from and to other proof systems and uses of LISA as a tool for program verification.
To improve usability and reduce the size of proofs, LISA makes use of an efficient normal form algorithm for propositional logic extended to first order logic. This algorithm is also the basis for a complete propositional proof-producing procedure implemented in LISA as a tactic.

LISA is still under active development, but already proposes an advanced proof writing DSL not entirely dissimilar to already existing interpreted languages in other assistants. LISA also allows defining arbitrary tactics in a simple way and has specialized error reporting.
The current embryo of set-theoretic development encompasses properties of relations and functions, and in particular Cantor's theorem has been successfully proven. 



{
\bibliographystyle{plainurl}
\raggedright
\bibliography{biblio.bib,vkuncak.bib}
}

\end{document}

%% file: macro.tex
\usepackage{url}
\usepackage{graphicx} 
\usepackage[utf8]{inputenc}
\usepackage{amsmath}
\usepackage{amsfonts}
\usepackage{amssymb}
\usepackage{listings}
\usepackage{amsthm}
\usepackage{bussproofs}
\usepackage[ruled,vlined]{algorithm2e}
\usepackage{array}
\usepackage{ifdraft}
\usepackage[dvipsnames]{xcolor}
\usepackage{dirtytalk}
\usepackage{float}
\usepackage{xspace}

\newtheorem{thm}{Theorem}

\theoremstyle{definition}

\newcommand{\schem}[1]{{'\hspace{-0.2em}#1}}
\newcommand{\OL}{\mathit{O{\kern-0.1em}L}}

\newcommand{\FOLalg}{F(OL)\textsuperscript{2}}
\newcommand{\FOLm}{\mathit{F{\kern-0.1em}({\kern-0.1em}O{\kern-0.1em}L{\kern-0.1em})\textsuperscript{2}}}

\newcommand{\ssimw}[1]{{{\kern-0.05em}/{\kern-0.08em}{#1}}}

\newcommand{\sime}{\sim_{{\kern-0.1em}E}}
\newcommand{\siml}{\sim_{{\kern-0.1em}L}}

\newcommand{\eqclass}[2]{{[#1]}_{#2}}
\newcommand{\eqclasssim}[2]{\eqclass{#1}{\sim_{{\kern-0.1em}#2}}}

\newcommand{\zf}{\textsf{ZF}\xspace}
\newcommand{\zfc}{\textsf{ZFC}\xspace}
\newcommand{\tg}{\textsf{TG}\xspace}

\newcommand{\simon}[1]{}
\newcommand{\viktor}[1]{}
\newcommand{\sankalp}[1]{}

\ifdraft{
    \linenumbers
    \renewcommand{\simon}[1]{\textcolor{blue}{[Simon: #1]}}
    \renewcommand{\viktor}[1]{\textcolor{ForestGreen}{[Viktor: #1]}}
    \renewcommand{\sankalp}[1]{\textcolor{WildStrawberry}{[Sankalp: #1]}}
}{}

\usepackage[nomap]{FiraMono} 

\definecolor{green}{rgb}{0, 0.6, 0}

\definecolor{comments}{RGB}{80,0,110}

\lstdefinelanguage{scala}{
    alsoletter={@,=,>},
    keywordstyle = {\color{blue}},
    keywordstyle = [2]{\color{blue}},
    commentstyle = \color{comments},
    morekeywords = [2]{abstract, case, class, def, do, Input, Output, then,
        else, extends, false, free, if, implicit, match,
        object, true, val, var, while, sealed, or,
        for, dependent, null, type, with, try, catch, finally,
        import, final, return, new, override, this, trait,
        private, public, protected, package, throw},
    sensitive = true, 
    numbers=left,
    stepnumber=1,
    morecomment = [l]{//},
    morecomment = [s]{/*}{*/},
    morestring = [b]",  
    otherkeywords = {;,<<,>>,++},
}

\lstdefinelanguage{lisa}{
    alsoletter={@,=,>},
    keywordstyle = {\color{blue}},
    keywordstyle = [2]{\color{blue}},
    keywordstyle = [3]{\color{green}},
    keywordstyle = [4]{\color{teal}},
    commentstyle = \color{comments},
    morekeywords = [2]{abstract, case, class, def, do, Input, Output, then,
        else, extends, false, free, if, implicit, match,
        object, true, val, var, while, sealed, or,
        for, dependent, null, type, with, try, catch, finally,
        import, final, return, new, override, this, trait,
        private, public, protected, package, throw},
    morekeywords = [3]{have, andThen, thenHave, Theorem, by, DEF, The, Lemma, subproof, assume},
    sensitive = true, 
    numbers=left,
    stepnumber=1,
    morecomment = [l]{//},
    morecomment = [s]{/*}{*/},
    morestring = [b]",  
    otherkeywords = {;,<<,>>,++},
}

\DeclareMathOperator{\pick}{pick}

\newcommand{\smartparagraph}[1]{\smallskip\noindent\textbf{#1.}\ }